\documentclass[]{spie}  

\usepackage{amsmath,amsfonts,amssymb}
\usepackage{graphicx}
\usepackage[colorlinks=true, allcolors=blue]{hyperref}
\newcommand{\specialcell}[2][c]{%
  \begin{tabular}[#1]{@{}c@{}}#2\end{tabular}}
\title{Classification of bad pixels of the Hawaii-2RG detector of the ASTROnomical NearInfraRed CAMera}

\author[a]{N.~A.~Maslennikova}
\author[a]{N.~I.~Shatsky}
\author[a]{A.~M.~Tatarnikov}
\affil[a]{Lomonosov Moscow State University, Sternberg Astronomical Institute, 13 Universitetsky pr., Moscow, 119234, Russia}

\authorinfo{Further author information: \\A.M.T.: E-mail: andrew@sai.msu.ru, Telephone: +7 495 939 1661\\  N.A.M.: E-mail: maslennikova.na16@physics.msu.ru, Telephone: +7 495 939 1661}

\begin{document} 
\maketitle

\begin{abstract}
ASTRONIRCAM is an infrared camera-spectrograph installed at the 2.5-meter telescope of the CMO SAI.  The instrument is equipped with the HAWAII-2RG array. A bad pixels classification of the ASTRONIRCAM detector is proposed. The classification is based on histograms of the difference of consecutive  non-destructive readouts of a flat field. Bad pixels are classified into 5 groups: hot (saturated on the first readout), warm (the signal accumulation rate is above the mean value by more than 5 standard deviations), cold (the rate is under the mean value by more than 5 standard deviations),  dead (no signal accumulation), and inverse (having a negative signal accumulation in the first readouts). Normal pixels of the ASTRONIRCAM detector account for 99.6\% of the total. We investigated the dependence between the amount of bad pixels and the number of cooldown cycles of the instrument. While hot pixels remain the same, the bad pixels of other types may migrate between groups. The number of pixels in each group stays roughly constant. We found that the mean and variance of the bad pixels amount in each group and the transitions between groups do not differ noticeably between normal or slow cooldowns.  
\end{abstract}


\section{Introduction}

ASTRONIRCAM \cite{nadjip2017,zh2020} is a camera and slit spectrograph for the near infrared installed at the 2.5 meter telescope of the Caucasian Mountain Observatory of the Sternberg Astronomical Institute of Lomonosov Moscow State University \cite{shatsky2020}. The instrument is equipped with the HAWAII-2RG $2048\times2048$ HgCdTe detector (hereafter referred to as the H2RG). The Leach  controller \cite{leach} allows multiple non-destructive readouts (NDR) of the detector during exposure. Operating in four reading channels, the entire array of $2048\times2048$ pixels is digitized within 3.646 seconds (this value is thus the minimum exposure time). In the photometric mode, only the central area of the $1024\times1024$ size is used.

The cryostat is cooled by the liquid nitrogen. The camera is a facility instrument of the telescope, so it is kept at the operating temperature 77~K permanently. Approximately twice a year, the cryostat is heated to ambient temperature, the vacuum in the cryostat is renewed and after that the camera is re-filled with the nitrogen. During this procedure the rate of temperature change should not exceed 1~K/min according to the specifications of the detector. The transition in the temperature of the camera from an ambient temperature to 77~K shall be called {\em ``cooldown''} hereafter.

This work aims to study and classify bad pixels of the ASTRONIRCAM camera detector. This classification is the basis for choosing the optimal camera cooldown mode and methods for correcting bad pixels when processing observations.

\section{Cooldowns}

\begin{figure} [ht]
   \begin{center}
   \begin{tabular}{c} 
   \includegraphics[height=8cm]{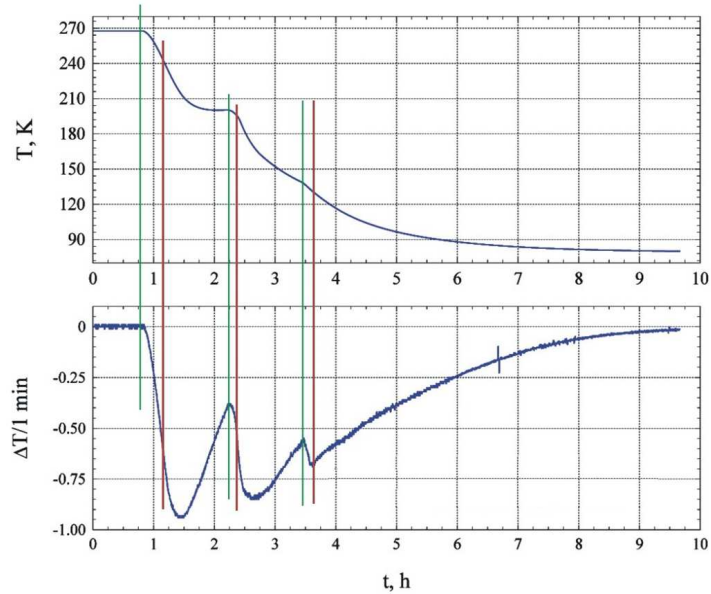}
   \end{tabular}
   \end{center}
   \caption[fig6] 
   { \label{fig:cooldown} 
Dependence of the cryostat temperature and the cooling rate on the time during a normal {\em cooldown}. Green and red lines denote the start and end of liquid nitrogen pouring, respectively.}
   \end{figure}

During {\em cooldown}, the liquid nitrogen is filled into the cryostat of the camera over several stages with a continuous control of the rate of temperature change (Fig.~\ref{fig:cooldown}). With a ``normal'' cryostat {\em cooldown}, when the cooling rate does not exceed 1~K/min, the active cooling phase lasts about 4~h. Still, in course of this process deformations of the hybrid structure arise so some contacts between the detector layers may brake off or recover. This can cause new bad pixels to appear or existing bad pixels to disappear. 

The transitions between different kinds of bad pixels after cooldowns are well observed in the ASTRONIRCAM detector and discussed in this paper. In order to check whether these changes depend on the rate of temperature change, a significantly slower {\em cooldown} was performed (date 07.07.2018 in Table~\ref{tab:tab1}). The rate of temperature change did not exceed 0.3~K/min, and the duration of the active phase increased to 12 hours.

\section{Previous bad pixel definitions}

There are various definitions of bad pixels in the literature. Sirianni et al. (2005) \cite{sirianni2005} consider pixels as hot if they generate ''very high'' dark current while Ingraham et al. (2014) \cite{ingraham2014} classify pixels into three groups: hot (the dark current is more than 1 electron per second; a typical pixel in their array has a dark current of less than 0.02 electrons per second), cold (having a response less than 15\% of the average as measured from the multi-filter pseudo-flat) and nonlinear (those demonstrating a strong nonlinear behavior at any level of exposure). Caron et al. (2013) \cite{caron2013} distinguish two groups: hot pixels (the response is above the mean in a dark current frame) and super-hot (saturated in the first reading of a flat field). One of the most detailed classifications is given in the work of  Hilbert (2012) \cite{hilbert2012} who divides bad pixels into 5 groups: Bad in Zeroth Read Pixels (a signal value is above 3 standard deviations (SD) in the zeroth read of a ramp), Dead Pixels (having small effective output), Unstable Pixels (with a signal changing all the time), Bad Reference Pixels, and Blobs (which appear because of particle hits into the channel selection mechanism). 

To determine whether the number of bad pixels changes from one {\em cooldown} to another and whether it depends on the temperature change rate during this procedure, we developed a new classification of bad pixels, since it turned out that the classification proposed by other authors does not describe adequately the features of the ASTRONIRCAM detector.

\begin{figure} [ht]
   \begin{center}
   \begin{tabular}{c} 
   \includegraphics[height=8cm]{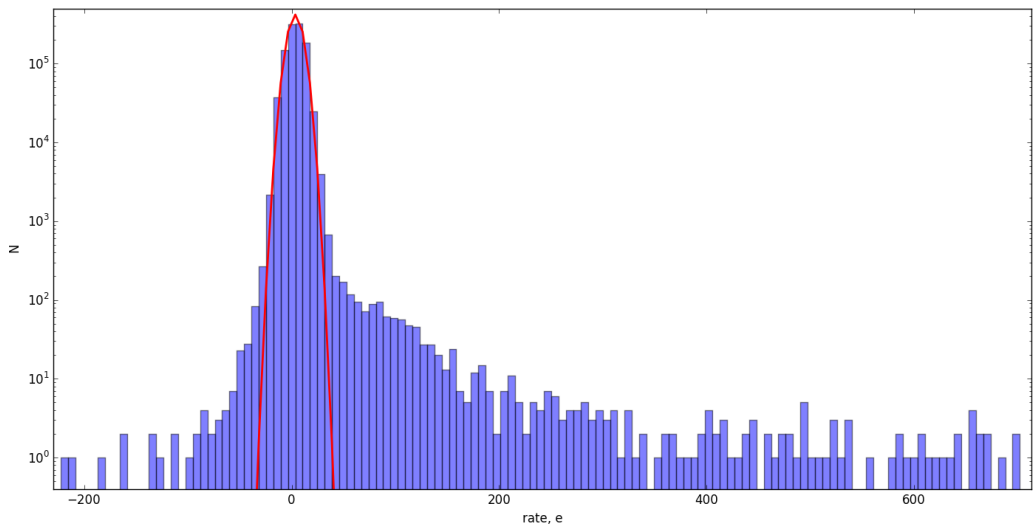}
   \end{tabular}
   \end{center}
   \caption[fig4] 
   { \label{fig:dark} 
Distribution of pixel signal accumulation rate in dark current frames}
   \end{figure} 
   
\section{Dark current}

Accumulation of the signal without incident light is the most traditional way to assess the pixel quality in CCD and especially infrared detectors. Fig.~\ref{fig:dark} demonstrates the distribution of pixels by the signal accumulation rate for a dark frame. It fits well with the normal distribution with a mean close to 0 and a SD corresponding to a readout noise. 

The detector H2RG has an extremely low mean value of dark current ($<0.1$ electrons per readout or $<0.03$ electrons per second) and a readout noise (RN) of around 12 electrons \cite{nadjip2017}. For the reliable identification of bad pixels of different types, we need the average signal value to be significantly greater than the readout noise. Only in this case, the ensembles of normal and dead pixels  will not overlap. On the H2RG dark current frames, normal pixels with small dark current and the dead pixels  that do not accumulate signal will not discern. Therefore, based on the dark frames, only one group of bad pixels could be distinguished -- these are warm pixels. About 90\% of them have a dark current  greater than 5~SD of the average dark current. So, a conventional analysis of dark frames cannot help to distinguish other groups to develop a detailed classification.

\section{Background and flat fields}

ASTRONIRCAM provides imaging capability in the standard photometric bands $J$, $H$, $K$ and $K\!s$. Distributions of the signal accumulation rate (see next section) for different filters and the night sky glow were constructed and approximated by the normal law. The maximum of the accumulation rate distribution is about of 2000 -- 3000 electrons per readout for the $H$ and $K$ bands which represent the high OH glow emission and a significant impact of thermal background, respectively. These values are high enough so the ranges of normal and dead+hot pixels distributions do not overlap. That is why we opted to use the background frames in these filters to classify bad pixels rather than dark frames.

The result of the pixels classification which we develop and discuss below does not depend on the particular photometric band which background frames were taken in. 

\section{Determination of signal accumulation rate} 

The H2RG detector has a zero signal level defined by the value of the potential at the signal lead of the photo-diode (a negative reverse bias). When the cell illumination takes place, the physical processes inside the p-n junction produce the positive growth of voltage at the signal lead proportionally to the intensity and exposure time of the illumination. Respectively,  the value of the initial p-n junction biasing effectively increases towards zero while the signal grows.

Infrared images are obtained in ASTRONIRCAM  by the "up-the-ramp" (or simply "Ramp") method \cite{fowler1991} implying periodic detector reading throughout the integration time (with the period of $\approx3.646$ seconds  for the readout of the full array). This way intermediate frames are obtained which are used to calculate the  dependence of a signal for each pixel on an NDR number. This dependence is approximated by a linear function which slope determines the signal accumulation rate, i.e. the value to be recorded in the final image.

\begin{figure} [ht]
   \begin{center}
   \begin{tabular}{c} 
   \includegraphics[height=8cm]{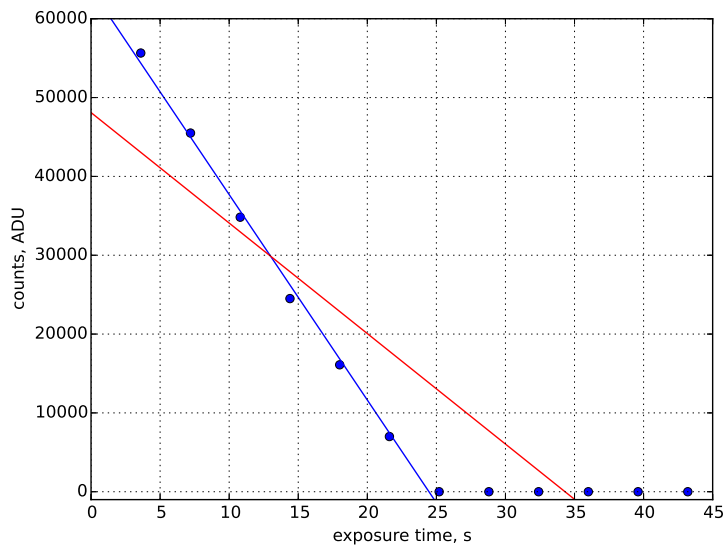}
   \end{tabular}
   \end{center}
   \caption[fig2] 
   { \label{fig:saturation} 
The dependence of the signal value on an exposure time for a pixel that is saturated during integration. The blue line is the approximating straight line that characterizes the real stimulus.  The red line is a formal regression slope computed from the whole integration data.}
   \end{figure}

Let us consider a pixel that is saturated during an exposure time (Fig.~\ref{fig:saturation}). At the beginning it quickly accumulates a signal. Therefore, the slope coefficient is large. When the pixel reaches the saturation limit, its signal value stops to change, so the approximating straight line  passes lower and the final regression  coefficient becomes smaller than the initial slope. In order to avoid such a biasing of results,  we chose to use  only the first two or three readouts in this work because most of the pixels do not saturate during this time. 

The master frames used to assess the signal accumulation rate for the bad pixel classification were obtained in the following way:

\begin{enumerate}

\item All frames are corrected for a master flat-field obtained by the 16~pix median filtering of a large number of individual flat field frames. This corrects some $\sim10$\% vignetting pattern within the field of view while biases only minorly the pixel thermal generation rates.
\item The frames of several integrations ($> 9$) obtained in the {\em dithering} mode while observing the night sky background are selected. The difference between the signal values in adjacent NDRs is calculated for each pixel for each flat field frame and converted to the number of electrons with $gain=2.2\rm{e^-/ADU}$\cite{nadjip2017}.
\item Median averaging is applied to the differences in order to get rid of the influence of cosmic particles and stars, the result is called the pixel's signal accumulation rate.
\end{enumerate}

\section{Classification by signal accumulation rate}

\begin{figure} [ht]
   \begin{center}
   \begin{tabular}{c} 
   \includegraphics[height=7cm]{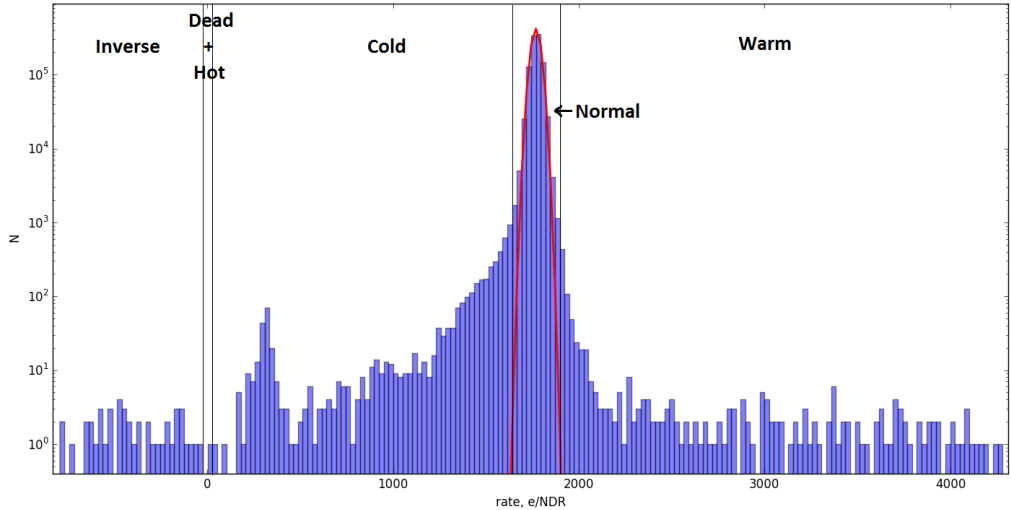}
   \end{tabular}
   \end{center}
   \caption[histrate] 
   { \label{fig:histrate} 
A pixel distribution by the accumulation rate in a typical master-frame. The histogram bin width is 1~SD.}
   \end{figure}

The distribution of pixels by the signal accumulation rate is shown in Fig.~\ref{fig:histrate}. We may expect that the rate distribution for ordinary (``good'') pixels is described by a normal law. In order to calculate its standard deviation SD, we removed bad pixels using a 5-sigma clipping procedure twice. For the normal distribution of $1K\!\times\!1K$ pixels, we can expect that there can be no more than 1 normal pixel out of this band which justifies the 5-sigma distinction used also throughout the rest of the paper. The rate SD is $\approx 40\rm{e^-/NDR}$ (2\% of the median rate).

\begin{figure} [ht]
   \begin{center}
   \begin{tabular}{c} 
   \includegraphics[height=10cm]{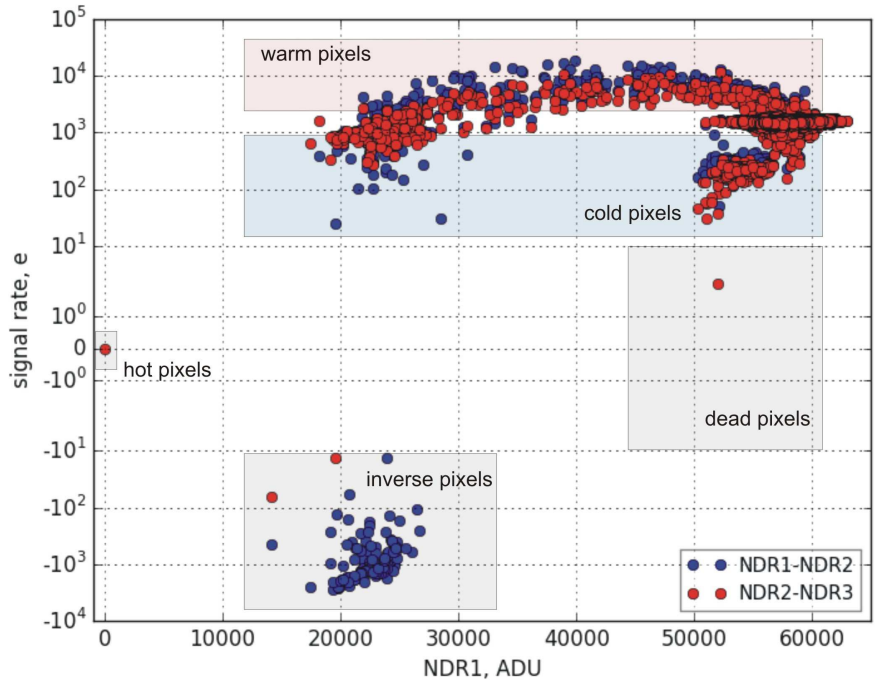}
   \end{tabular}
   \end{center}
   \caption[rate-vs-ndr] 
   { \label{fig:rate_ndr1} 
Diagram of the dependence of the signal accumulation rate on the signal value in the 1-st NDR}
   \end{figure} 

To the right and to the left of the distribution core we can see the prominent excesses of pixels with higher and lower (and even negative) rates of accumulation than those of the normal core. The practice of the detector data reduction forced us to consider the pixels variety not only by rates but also by their initial reset charge and apply them separately to the first two and subsequent NDRs.

In Fig.~\ref{fig:rate_ndr1} the signal accumulation rate of pixels within the photometric mode field of view is plotted against the signal in the first NDR. Blue points here denote the rate determined from the NDR1$-$NDR2 difference while red ones show the rate between the second and third NDRs. Points are clearly attracted to a few regions or curves. The vast majority occupy the short horizontal line of pixels having the charge between some 50 and 63 thousand ADU and represent the normal core of the distribution shown in Fig.~\ref{fig:histrate}. Among others are those in a compact group of the reduced accumulation rate and following a wide ``arch'' spreading across the regions of excessive to moderately reduced rates and finally reaching the significant deficit of the pixel charge at the start of accumulation. Finally, we see a well isolated group of initially undercharged pixels possessing the abnormal negative difference between the 1st and 2nd readouts.

A cloud of blue points at the top of the Fig.~\ref{fig:rate_ndr1} clearly seems up-shifted relative to the respective red dots population. A dedicated comparison of the relative position of blue and red dots for each pixel witnesses that pixels within the bands of excessive and reduced rates in the upper part of the diagram indeed possess the higher rates between NDR1 and NDR2 than within the adjacent time span. The rates difference depends also on the NDR1 level: the lower is this initial charge the higher is the difference. The subsequent NDRs demonstrate for these pixels the discharge rate characteristic of the normal pixels (see below).

The ``arch'' crosses again the rate level of the core (normal) distribution at a significantly lower NRD1 value, between 20 and some 35 thousand ADU. Thus, there is a considerable number of pixels which mimic the normal majority but have a halved initial charge. Although it is clear that these pixels have a different physical behaviour, in what follows we omit this difference due to their normal accumulation through the rest of the ramp.

Summarizing this brief analysis of pixels distribution in a quasi-3D space of rates, their differences between NDR1-NDR2 and NDR2-NDR3 (and so on) and the initial signal, and refraining from the further hypothesizing on the physical nature of related groups, we can propose the following phenomenological classification of pixels in the ASTRONIRCAM detector:

\begin{description}
\item[Normal] pixels are pixels which signal accumulation rate differs from the mean by less than 5~SD. Most of them are located in the region ($NDR1=52k-63$k, signal rate$\approx 2000$) in Fig.~\ref{fig:rate_ndr1} but include also the undercharged group of pixels with the same accumulation rate.

\item[Cold] pixels accumulate signal slower than 5~SD below the mean rate, but the accumulation rate is greater than the detector readout noise. They are located mainly in the region (NDR1=50k--60k, signal rate $<2000$) in Fig.~\ref{fig:rate_ndr1}. Among the cold pixels, a small but constant in composition group stands out which could be called ``super cold''. It is clearly visible in Fig.~\ref{fig:histrate} just to the right of the rate 0 and in Fig.~\ref{fig:rate_ndr1} as well. However, from the point of view of a subsequent correction procedure, these pixels do not differ from other cold pixels, so we will not distinguish this group in what follows.

\item[Warm] pixels accumulate signal faster (and sometimes far faster) than the mean rate plus 5~SD. As it follows from the next section data and a primitive logic, these pixels demonstrate the true signal plus a significant dark current.  Warm pixels are located mainly in a wide arched area above the normal pixels level in Fig.~\ref{fig:rate_ndr1}. However, some of them fall into the region of the diagram, which should formally belong to cold pixels at a low NDR1 value. It should be stressed here that ``normal warm'' pixels are those generating excessive dark current signal, see below.

\item[Dead] pixels do not accumulate a signal at all, but due to the existence of readout noise, the signal in them will be in the range of  (-RN, RN).

\item[Hot] pixels have a saturated signal at first readout. Thus, the accumulation rate of a hot pixel is also 0, but unlike dead ones, the signal value for hot pixels in each intermediate reading is exactly 0.

\item[Inverse] pixels have a negative signal accumulation rate between 1 and 2 NDRs. These pixels represent the most isolated group in Fig.~\ref{fig:rate_ndr1} but all but two of them have a positive (i.e. normal) accumulation rate in subsequent readings of a ramp (two pixels may possess negative rate over three readouts).

\end{description}

\section{Bad pixels behavior and correction}

Let us consider in more detail the behavior of pixels from different groups during signal accumulation.

Hot and dead pixels do not accumulate signal, so information in these pixels is completely lost. During the correction procedure, we replace the signal in them with the average signal value in neighboring pixels.

\begin{figure}[ht]
\begin{minipage}[h]{0.5\linewidth}
\center{\includegraphics[width=1\linewidth]{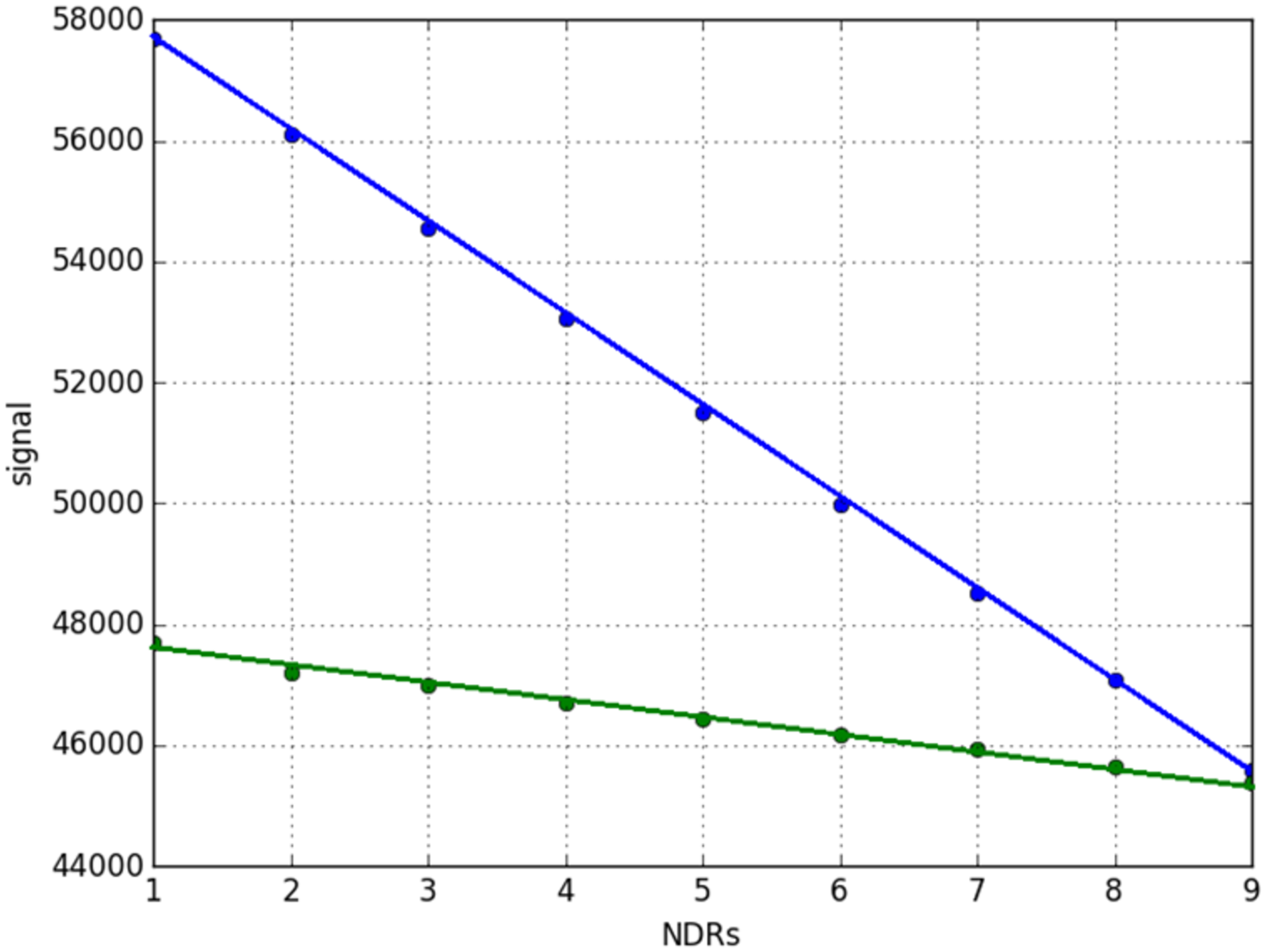} \\a)}
\end{minipage}
\hfill
\begin{minipage}[h]{0.5\linewidth}
\center{\includegraphics[width=1\linewidth]{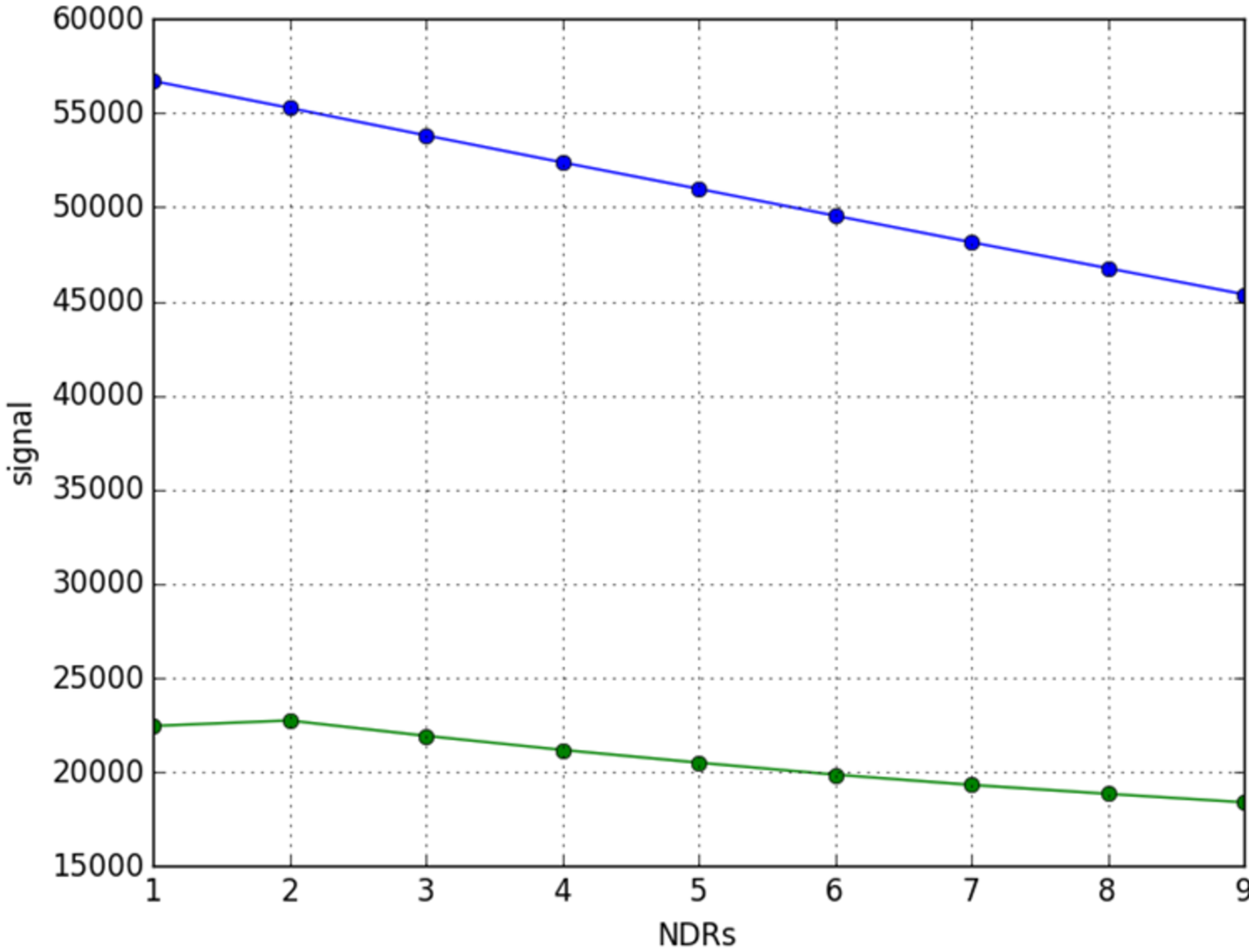} \\b)}
\end{minipage}
\vfill
\begin{minipage}[h]{0.5\linewidth}
\center{\includegraphics[width=1\linewidth]{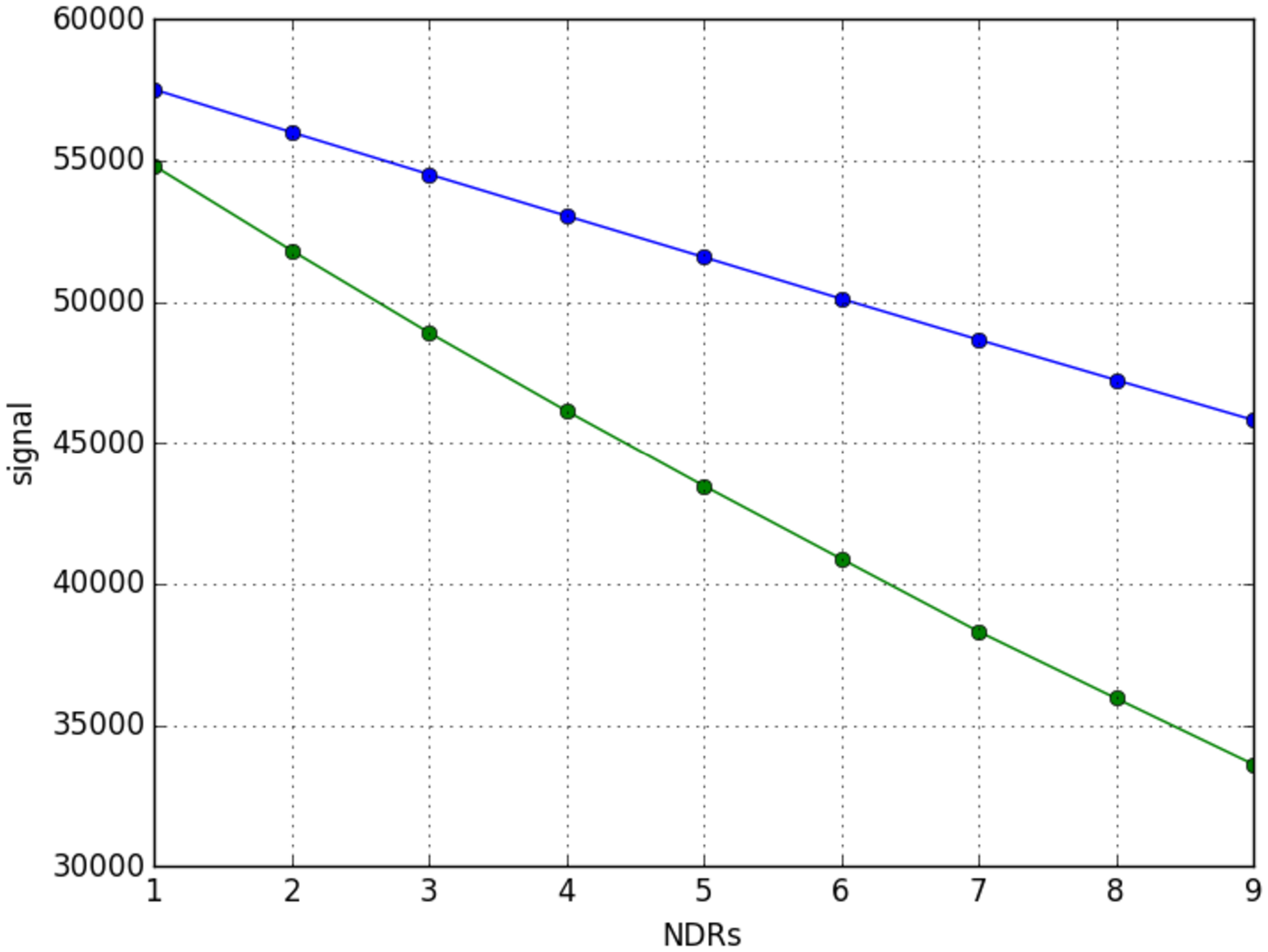} \\c)}
\end{minipage}
\hfill
\begin{minipage}[h]{0.5\linewidth}
\center{\includegraphics[width=1\linewidth]{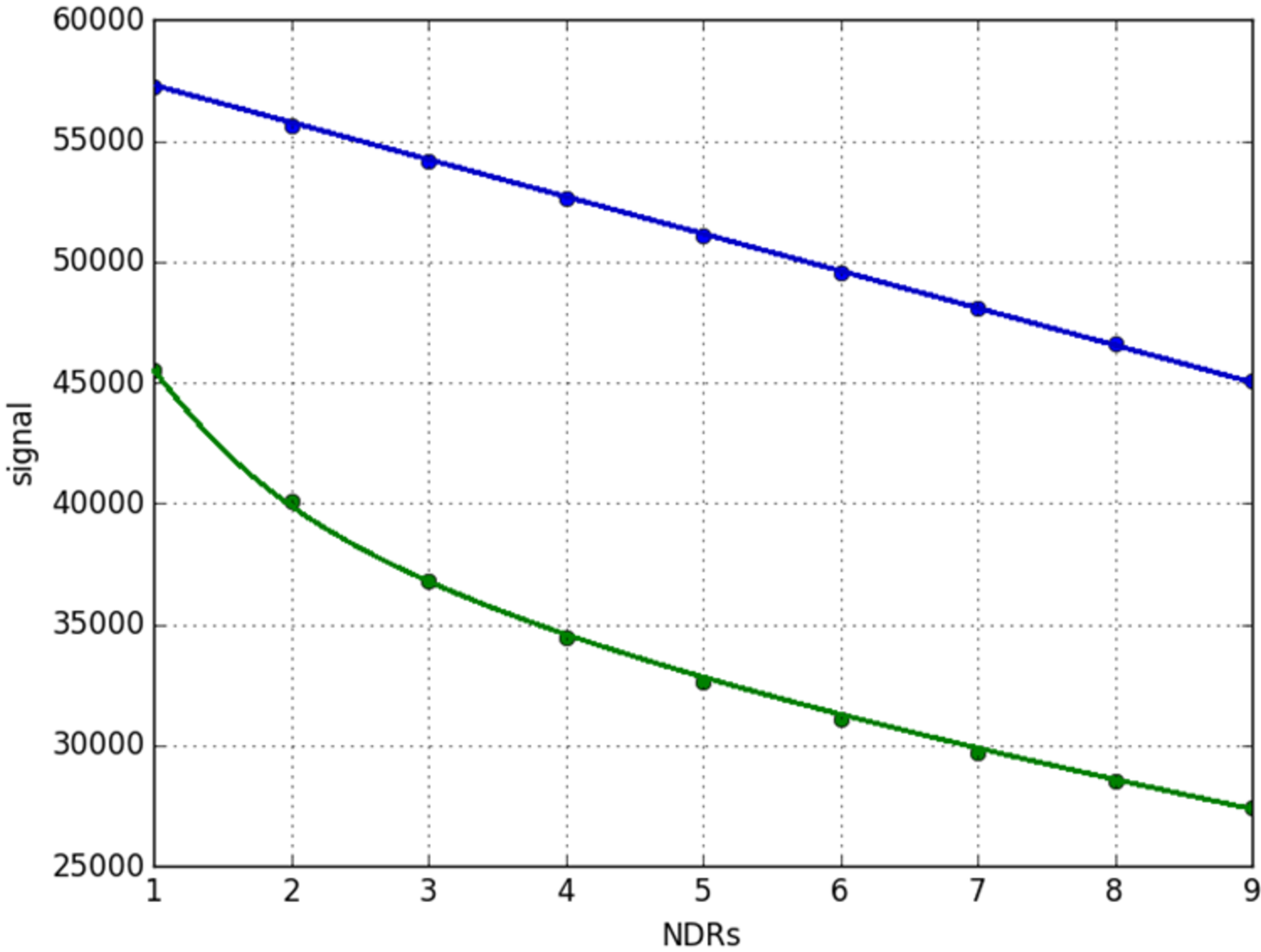} \\d)}
\end{minipage}
\caption[fig7]
{\label{fig:bad_curves}
Signal accumulation curves plotted for normal (a blue line) and bad pixels (a green line): a) a cold pixel, b) an inverse pixel, c) a linear warm pixel, d) a nonlinear warm pixel. The abscissa axis shows the number of non-destructive readouts, the ordinate axis shows the accumulated signal in ADU.}
\end{figure}

Cold pixels do not differ from normal ones in the shape of the signal accumulation curve, only in the accumulation rate (see Fig.~\ref{fig:bad_curves}a). Therefore, the signal value in  cold pixels can be corrected during processing, information is not lost in them.

The signal accumulation curve of warm pixels can be almost linear (Fig.~\ref{fig:bad_curves}c) or  strongly nonlinear (Fig.~\ref{fig:bad_curves}d). If a warm pixel has a strong nonlinearity then it cannot be corrected, the information in it is completely lost. If it is nearly linear then it can be corrected. 

An inverse pixel (Fig.~\ref{fig:bad_curves}b) has a negative signal accumulation rate between the first and second readouts. It then behaves like a cold pixel. But unlike a cold pixel, it can be corrected only with a large number of readouts; in bright objects accumulations (i.e., with a small number of readouts) it has to be invalidated.

\section{Bad pixel transitions}
The detector is cooled with liquid nitrogen; its operating temperature varies within 77.4--78.4~K (ASTRONIRCAM temperature is not stabilized). As it is mentioned above, when the instrument is cooled from an ambient to the operating temperature bad pixels may occur or become back normal (when restoring sufficient contact), or move from group to group.

\begin{figure} [ht]
   \begin{center}
   \begin{tabular}{c} 
   \includegraphics[height=10cm]{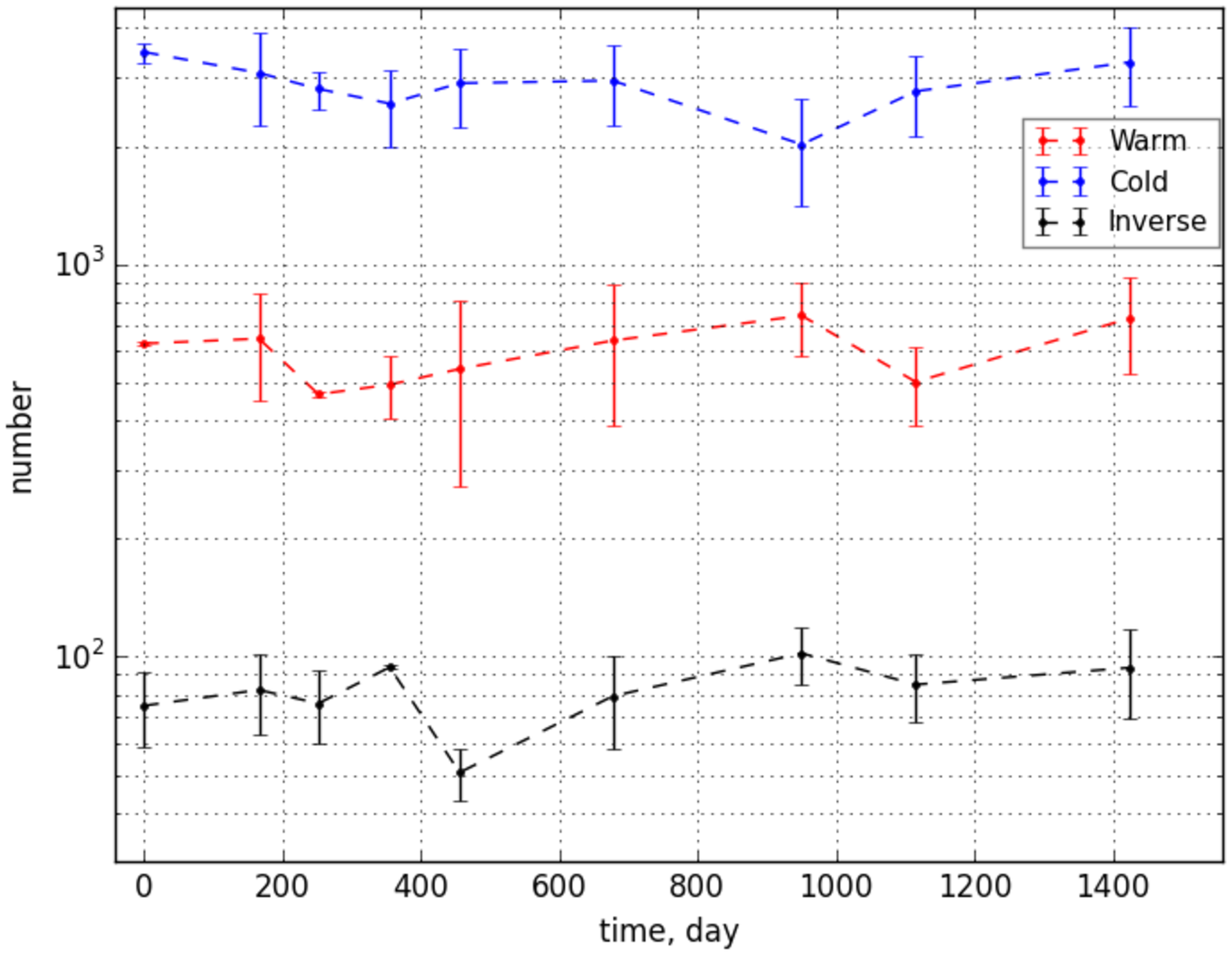}
   \end{tabular}
   \end{center}
   \caption[figres] 
   { \label{fig:bads_evolutions} 
The change in the number of bad pixels over time}
   \end{figure}

We examined a large number of flat field frames obtained in the filter $H$ between {\em cooldowns} in 2015--2020. Normal pixels of the ASTRONIRCAM detector account for 99.6\% of the total. The number of pixels in each group of bad pixels is given in Table~\ref{tab:tab1}.

The most interesting phenomenon is the transition of pixels between groups after a cooldown was performed (see examples in Table~\ref{tab:tab2}). For a quantitative characterization of transitions we performed the following procedure on the basis of $H$- and $K$-band flat fields (which are not observed every night):

\begin{enumerate}
\item The last flats observation night is taken before the instrument maintenance period and the first night after a subsequent cooldown.
\item The pixels are classified as described above for both nights.
\item The bad pixel maps are compared and the pixels which have changed their class are identified.
\end{enumerate}

Let us consider the pixels transition across the cooldown dated, say, 20160522 (see Table~\ref{tab:tab2}). The respective pre- and post-cooldown flats were observed on 20160422 and 20160602, respectively, so pixels of these nights are to be classified. The number of warm pixels before the cooldown was 636 (see the first cell of the table). Out of these, 441 remained warm (notice the record ``W-441'') while 4 pixels became cold (``C-4''). There were no transitions to inverse or dead pixels, so ``I-0, D-0'' is indicated, while rest of the pixels became normal (``N-191''). This way the rest of the table is filled as well.

Only hot pixel locations are exactly constant (the same 6 pixels) and do not change with {\em cooldowns}. Most of bad pixels don't change their group over time as well. Warm and cold pixels when leaving their group join mostly the normal pixels and back, too. It is also seen that there is no noticeable difference of transitions between groups happening across dates of normal or slow {\em cooldowns}.

The most frequent transition between groups is, naturally, warm and cold pixels transition to normal or vice versa. To determine whether the pixels really change their type, or transitions occurs only statistically (due to a small shift in the border between groups and a insignificant change in the rate of pixel signal accumulation near the border), we examined the transitions between groups in more detail.

The transition between warm and normal pixels (Table~\ref{tab:tab3}): within the range $\pm$SD from the border separating these two types of pixels there are approximately 40$\%$ of such pixels, within the $\pm$2~SD range --- about 65$\%$. Such a large number cannot be explained by random fluctuations for a normal distribution of $10^6$ pixels. Thus, the $60\%$ of warm pixels that turn into normal pixels are actually seen to change their physical behavior. Meanwhile, since the statistics of the bad pixel rates may not obey the normal law, the reality and physical sense of these transitions require further proof with a far more detailed analysis than presented in this study.

The transition between cold and normal pixels: from the right part of Table~\ref{tab:tab3} we can see that cold pixels have the same percentage as warm pixels in these ranges, they also actually change their physical behavior when they turn into normal pixels.

The Fig.~\ref{fig:bads_evolutions} shows the average number of warm, cold and inverse pixels observed between different {\em cooldowns}. The first six and last two points of each curve correspond to normal {\em cooldowns}, at which a relatively high cooling rate was maintained (less or about 1K/min, see the section on cooldowns above). We can see that the number of bad pixels in the detector is independent of the detector cooling rate during {\em cooldowns}. We also do not observe any noticeable trend in the number of bad pixels of our detector over 5 years of exploitation.

\section{Conclusion}

We proposed a classification of bad pixels of detectors operating in the non-destructive readout mode. The classification does not attempt to provide the physical explanation of the pixels behaviour and uses the distribution of pixels by the difference of signals accumulated in the first two readouts during observations of flat fields (night sky background). Bad pixels are formally classified into 5 groups: {\em warm} (the signal is more than 5~SD above the mean), {\em cold} (a signal is less than 5~SD below the mean, but more than 0), {\em hot} (those already saturated in the first readout), {\em dead} (no signal accumulation), and {\em inverse} (having a negative signal in the first readouts). The number of normal pixels in the central photometric area of the ASTRONIRCAM detector is 99.6\%. There are about 4,000 bad pixels there. Around 100 of them cannot be corrected, which is about 0.01\% of the total number of pixels (we don not take into account the two defective detector rows here).

The migration of bad pixels between groups happening after {\em cooldowns} (the process of filling a warm cryostat with liquid nitrogen) was studied according to adopted formal criteria. Most of pixels reside in the same group. Hot pixels are represented by the same 6 pixels. The number of dead pixels varies but it is always less than 10. They may remain dead or become cold or inverse. There are about 100 inverse pixels. Sometimes they can turn into groups of normal, cold or warm pixels. The number of warm pixels is 400--700. Cold pixels are the largest group of bad pixels (2000--4000). More than 80\% of cold and 50\% of warm pixels remain in their groups. The remaining pixels mostly turn into normal pixels.

We studied the dependence of the total number of bad pixels on the number of {\em cooldowns} and the cooling rate of the cryostat during this procedure. It turned out that the number of bad pixels after a {\em cooldown} remains statistically at the same level as was during the instrument operation period. It does not depend on the cooling rate of the camera either.

\acknowledgements
NS (results presentation) and AT (reduction algorithms) acknowledge the partial support by the Russian Science Foundation grant 17-12-01241. This research has been supported by the Interdisciplinary Scientific and Educational School of Moscow University ''Fundamental and Applied Space Research''.


\begin{table}[ht]
\caption{The number of pixels in each group of bad pixels. The top row contains the dates of successive {\em cooldowns} between which numbers of bad pixels were computed}
\label{tab:tab1}
\begin{center}
\begin{tabular}{|l|l|l|l|l|l|l|l|}
\hline
  \specialcell{Dates of\\{\em cooldown}}& \specialcell{2015-11-01 \\ 2016-05-22}  & \specialcell{2016-05-22 \\ 2016-09-28} & \specialcell{2016-09-28 \\ 2016-11-14} & \specialcell{2016-11-14 \\ 2017-04-20} & \specialcell{2017-04-20 \\ 2017-05-30} & \specialcell{2017-05-30 \\ 2018-07-07} & \specialcell{2018-07-07  \\ 2018-12-02}\\
\hline
{\em Dead} & 1 & 2 & 1 & 2 & 1 & 2 & 3 \\
\hline
{\em Hot} & 6 &  6 &  6 &  6 &  6 &  6 &  6 \\
\hline
{\em Cold} & $3146\pm366$& $2896\pm832$ & $3058\pm429$ & $2821\pm637$ & $3042\pm623$ & $2991\pm694$ & $2025\pm568$ \\
\hline
{\em Warm} & $522\pm115$ & $578\pm234$ & $527\pm83$ & $595\pm121$ & $553\pm234$ & $687\pm331$ & $717\pm148$ \\
\hline
{\em Inverse} & $75\pm16$ & $82\pm19$& $76\pm16$& $94\pm1$& $51\pm8$ & $79\pm21$& $112\pm15$ \\
\hline
\specialcell{Total\\number} & 3750& 3564 & 3668 & 3518 & 3653 & 3765 & 2863 \\
\hline
\end{tabular}
\end{center}
\end{table}

\begin{table}[t]
\caption{Change of bad pixel types. The first column shows the pixel groups before the restore. 
The next columns show the number of pixels and the group to which they were transferred during a cooldown
of the date indicated in the first row.}
\label{tab:tab2}
\begin{center}
\begin{tabular}{|c|c|c|c|c|c|c|}
\hline
Dates & 2016-05-22 & 2016-09-28 & 2016-11-14 & 2017-04-20 & 2017-05-30 & 2018-07-07 \\
\hline
Hot (H) & H – 6 & H – 6 & H – 6 & H – 6 & H – 6 & H – 6 \\
\hline
Warm (W) & \specialcell{W – 441 \\
C – 4 \\
D – 0 \\
I – 0 \\
N – 191 \\
}  & \specialcell{W – 385 \\
C – 4\\
D – 0 \\
I – 0\\
N – 413\\
} & \specialcell{W – 372\\
C – 1\\
D – 0 \\
I – 0\\
N – 84\\
} & \specialcell{W – 266\\
C – 5\\
D – 0 \\
I – 0\\
N – 132\\
} & \specialcell{W – 262\\
C – 2\\
D – 0 \\
I – 0\\
N – 71 \\
} & \specialcell{W – 374\\
C – 2\\
D – 0 \\
I – 1\\
N – 326\\
}  \\ 
\hline
Cold (C) & \specialcell{W – 1 \\
C – 2803 \\
D – 0 \\
I – 0 \\
N – 472 \\
}  & \specialcell{W – 1 \\
C – 2892\\
D – 1 \\
I – 0\\
N – 814\\
} & \specialcell{W – 2\\
C – 2339\\
D – 1 \\
I – 0\\
N – 153\\
} & \specialcell{W – 3\\
C – 1851\\
D – 0 \\
I – 0\\
N – 142\\
} & \specialcell{W – 2\\
C – 1940\\
D – 0 \\
I – 0\\
N – 360 \\
} & \specialcell{W – 10\\
C – 1910\\
D – 2 \\
I – 0\\
N – 1321\\
}  \\ 
\hline
Dead (D) & \specialcell{W – 0 \\
C – 1 \\
D – 0 \\
I – 0 \\
N – 1 \\
}  & \specialcell{W – 0 \\
C – 0\\
D – 0\\
I – 0\\
N – 1\\
} & \specialcell{W – 0\\
C – 1\\
D – 1 \\
I – 0\\
N – 0\\
} & \specialcell{W – 0\\
C – 1\\
D – 0 \\
I – 0\\
N – 1\\
} & \specialcell{W – 0\\
C – 0\\
D – 1 \\
I – 0\\
N – 0 \\
} & \specialcell{W – 0\\
C – 0\\
D – 1 \\
I – 0\\
N – 0\\
}  \\ 
\hline
Inverse (I) & \specialcell{W – 0 \\
C – 0\\
D – 0 \\
I – 89\\
N – 2\\
} & \specialcell{W – 0\\
C – 0\\
D – 0 \\
I – 55\\
N – 2\\
} & \specialcell{W – 0\\
C – 0\\
D – 0\\
I – 87\\
N – 5\\
} & \specialcell{W – 0\\
C – 0\\
D – 0 \\
I – 40\\
N – 55 \\
} & \specialcell{W – 0\\
C – 0\\
D – 0 \\
I – 57\\
N – 0\\
}  & \specialcell{W – 0\\
C – 0\\
D – 0 \\
I – 87\\
N – 2\\
}  \\ 
\hline
Normal (N) & \specialcell{W – 150 \\
C – 356 \\
D – 1 \\
I – 0 \\
}  & \specialcell{W – 90 \\
C – 211\\
D –  1\\
I – 5\\
}  & \specialcell{W – 210 \\
C – 802\\
D –  0\\
I – 6\\
} & \specialcell{W – 98\\
C – 716\\
D – 1 \\
I – 0\\
} & \specialcell{W – 91\\
C – 301\\
D – 0 \\
I – 28\\
} & \specialcell{W – 195\\
C – 137\\
D – 0 \\
I – 36\\
}  \\ 
\hline
\end{tabular}
\end{center}
\end{table}

\begin{table}[h]
\caption{The number of normal pixels that turn to the warm or cold pixels group and vice versa. The columns indicate the number of pixels lying in the given range from all that changed the normal group to the warm (cold) group or vice versa. }
\label{tab:tab3}
\begin{center}
\begin{tabular}{|c|c|c|c|c|}
\hline
 Transition & \multicolumn{2}{|c|}{Normal$\rightarrow$Warm} & \multicolumn{2}{|c|}{Normal$\rightarrow$Cold}\\ \hline
  \specialcell{Date} & \specialcell{$4\ SD<x<6\ SD$}  & \specialcell{$3\ SD<x<7\ SD$ } & \specialcell{$4\ SD<x<6\ SD$}  & \specialcell{$3\ SD<x<7\ SD$ }\\ 
\hline
2016-11-14 & 115/210\ (54\%) & 175/210\ (83\%) 
& 433/802\ (54\%) & 699/802\ (87\%)\\ 
\hline
2017-04-20 & 30/98\ (31\%) & 61/98\ (62\%) 
& 364/716\ (51\%) & 572/716\ (80\%)\\ 
\hline
2017-05-30 & 39/91\ (40\%) & 57/91\ (63\%) 
& 92/301\ (31\%) & 195/301\ (65\%)\\ 
\hline
2018-07-07 & 60/195\ (31\%) & 96/195\ (49\%)
& 78/137\ (57\%) & 106/137\ (77\%)\\ 
\hline
 Transition & \multicolumn{2}{|c|}{Warm$\rightarrow$Normal} & \multicolumn{2}{|c|}{Cold$\rightarrow$Normal}\\ \hline
  \specialcell{Date} & \specialcell{$4\ SD<x<6\ SD$}  & \specialcell{$3\ SD<x<7\ SD$ } & \specialcell{$4\ SD<x<6\ SD$}  & \specialcell{$3\ SD<x<7\ SD$ }\\ 
\hline
2016-11-14 & 40/84\ (48\%) & 58/84\ (69\%)
& 80/153\ (52\%) & 133/153\ (87\%)\\ 
\hline
2017-04-20 & 13/132\ (10\%) & 37/132\ (28\%)
& 63/142\ (44\%) & 114/142\ (80\%)\\ 
\hline
2017-05-30 & 36/71\ (51\%) & 49/71\ (69\%)
& 44/360\ (12\%) & 133/360\ (37\%)\\ 
\hline
2018-07-07 & 224/326\ (69\%) & 304/326\ (93\%)
& 445/1321\ (34\%) & 823/1321\ (62\%)\\ 
\hline
\end{tabular}
\end{center}
\end{table}

\end{document}